\documentclass[prl,twocolumn,showpacs,byrevtex,showkeys]{revtex4}%
\usepackage{amsfonts}
\usepackage{amsmath}
\usepackage{amssymb}
\usepackage{graphicx}
\usepackage{hyperref}
\providecommand{\U}[1]{\protect\rule{.1in}{.1in}}
\def\ket#1{\vert#1\rangle}

\let\originalleft\left
\let\originalright\right
\renewcommand{\left}{\mathopen{}\mathclose\bgroup\originalleft}
\renewcommand{\right}{\aftergroup\egroup\originalright}

\begin{document}
\preprint{ }
\title[ ]{Quantum state cloning using Deutschian closed timelike curves}
\author{Todd A. Brun}
\affiliation{Ming Hsieh Department of Electrical Engineering, University of Southern
California, Los Angeles, California 90089, USA}
\author{Mark M. Wilde}
\affiliation{Department of Physics and Astronomy, Center for Computation and Technology,
Louisiana State University, Baton Rouge, Louisiana 70803, USA}
\author{Andreas Winter}
\affiliation{ICREA \& F\'{\i}sica Te\`{o}rica: Informaci\'{o} i Fenomens Qu\`{a}ntics,
Universitat Aut\`{o}noma de Barcelona, ES-08193 Bellaterra (Barcelona), Spain}
\affiliation{School of Mathematics, University of Bristol, Bristol BS8 1TW, United Kingdom}

\begin{abstract}
We show that it is possible to clone quantum states to arbitrary accuracy in
the presence of a Deutschian closed timelike curve (D-CTC), with a fidelity
converging to one in the limit as the dimension of the CTC\ system becomes
large---thus resolving an open conjecture from [Brun \textit{et al}., Physical
Review Letters \textbf{102}, 210402 (2009)]. This result follows from a
D-CTC-assisted scheme for producing perfect clones of a quantum state prepared
in a known eigenbasis, and the fact that one can reconstruct an approximation
of a quantum state from empirical estimates of the probabilities of
an informationally-complete measurement. Our results imply more generally
that every continuous, but otherwise arbitrarily non-linear map
from states to states can be implemented to arbitrary accuracy with D-CTCs.
Furthermore, our results show that Deutsch's model for CTCs is in fact a classical model, in the sense  that two arbitrary, distinct density operators are perfectly distinguishable (in the limit of a large CTC system); hence, in this model quantum mechanics becomes a classical theory in which each density operator is a distinct point in a classical phase space.
\end{abstract}
\volumeyear{2013}
\volumenumber{ }
\issuenumber{ }
\eid{ }
\date{\today}
\startpage{1}
\endpage{10}
\maketitle

The possible existence of closed timelike curves (CTCs)\ in certain exotic
spacetime geometries \cite{RevModPhys.21.447,PhysRevLett.66.1126,B80}\ has
sparked a significant amount of research regarding their ramifications for
computation \cite{fpl2003brun,PhysRevA.70.032309,ScottAaronson02082009}\ and
information processing \cite{PhysRevLett.102.210402,PhysRevLett.110.060501}.
One of the well known models for CTCs is due to Deutsch \cite{PhysRevD.44.3197},
who had the insight to abstract away much of the space-time geometric
details and use the tools of quantum information to address physical questions
about causality paradoxes. One consequence is that quantum computers
with access to \textquotedblleft Deutschian\textquotedblright\ CTCs (D-CTCs)
would be
able to answer any computational decision problem in PSPACE
\cite{ScottAaronson02082009}, a powerful complexity class containing the
well-known class NP, for example. Also, quantum information processors with
access to D-CTCs could distinguish non-orthogonal states
perfectly \cite{PhysRevLett.102.210402}, thus leading to the strongest
violation of the uncertainty principle that one could imagine. From the
perspective of Aaronson \cite{A05,Aaronson08112005}, we might take these results
to be complexity- and information-theoretic evidence against the existence of
CTCs that behave according to Deutsch's model.

In order to avoid ``grandfather-like'' paradoxes, Deutsch's model imposes a boundary condition, in which the density operator of the CTC\ system before it has interacted with a chronology-respecting system should be equal to the density operator of the CTC\ system after it interacts. More formally, let $\rho_{S}$ denote the state of the chronology-respecting system and let $\sigma_{C}$ denote the state of the CTC\ system before a unitary interaction $U_{SC}$ (acting on systems $S$ and $C$) takes place. The first assumption of Deutsch's model is that the state of the chronology-respecting system $S$ and the chronology-violating system $C$ is a tensor-product state, since presumably they have not interacted before the CTC\ system comes into existence.  Furthermore, Deutsch's model imposes the following self-consistency condition:%
\begin{equation}
\sigma_{C}=\Phi_{\rho}\left(  \sigma_{C}\right)  \equiv\text{Tr}_{S}\left\{
U_{SC}\left(  \rho_{S}\otimes\sigma_{C}\right)  U_{SC}^{\dag}\right\}  , \label{eq:fixed-point}
\end{equation}
so that potential grandfather paradoxes can be avoided.
Computationally, one can take the view that nature is finding a
fixed point of the map $\Phi_{\rho}$ \cite{PhysRevD.44.3197,ScottAaronson02082009},
which depends on the state $\rho_{S}$ of the chronology-respecting
system.  The chronology-respecting system's state evolves by
\[
\rho_{S} \rightarrow \rho_{\rm out} = \text{Tr}_{C}\left\{
U_{SC}\left(  \rho_{S}\otimes\sigma_{C}\right)  U_{SC}^{\dag}\right\}  ,
\]
where the partial trace is over the CTC\ system.  Since $\sigma_C$ depends
on $\rho_{S}$, such an evolution is non-linear and as a result
is a non-standard quantum evolution.

In developing the above consistency condition, Deutsch explicitly
assumed that density operators are the fundamental object
characterizing quantum systems, and, under this assumption, Deutsch's
model does not lead to any of the classical time-travel paradoxes \cite{PhysRevD.44.3197}.
If the density operator is viewed as a statistical ensemble or as a
state of knowledge, then Deutsch's consistency condition becomes
problematic and could conceal underlying paradoxes \cite{PhysRevD.44.3197,WB12}.

Since quantum processors with access to D-CTCs can perfectly
distinguish pure quantum states \cite{PhysRevLett.102.210402}, one might
conclude that such D-CTC-assisted processors could also
approximately clone any pure quantum state, in violation of the celebrated
no-cloning theorem \cite{WZ82,D82}. In fact, Deutsch suggested
that quantum cloning should be possible when one has access to
D-CTCs behaving according to (\ref{eq:fixed-point}) \cite{PhysRevD.44.3197},
and Brun \textit{et
al}.~conjectured that
``a [D-CTC-assisted] party can construct a universal cloner
with fidelity approaching one, at the cost of increasing the available
dimensions in ancillary and CTC resources'' \cite{PhysRevLett.102.210402}.
Indeed, a simple
idea for building an approximate cloner would be to discretize a given
finite-dimensional Hilbert space, by casting an $\varepsilon$-net over all of
the pure states in it, such that any state in the Hilbert space is
$\varepsilon$-close in trace distance to a state in the $\varepsilon$-net.
(Simple arguments for the size of such $\varepsilon$-nets are well known
\cite{HLSW04}.) One would then construct a unitary for perfectly
distinguishing states in the $\varepsilon$-net, according to the procedure
given in \cite{PhysRevLett.102.210402}, and produce clones according to the
classical outcome of the distinguishing device. States in the $\varepsilon
$-net would be cloned perfectly, while the hope is that states that are not in
the $\varepsilon$-net would be identified with the closest state in the
$\varepsilon$-net.

An approach similar to this was pursued in \cite{AMRM12},
and the numerical evidence given there suggests that such an approach should
work in general. However, it is well known (and perhaps obvious)\ that there
are continuity issues with D-CTCs
\cite{PhysRevD.44.3197,ScottAaronson02082009,PhysRevD.81.087501}, so that one
cannot easily appeal to continuity in order to develop this argument in
greater detail.

In this paper, we give an approach to quantum state cloning with D-CTCs that is
conceptually different from the aforementioned one, and it is also
significantly simpler and thus more appealing. We show how to clone any
quantum state, such that the fidelity of each clone approaches one as the
dimension of the assisting D-CTC system becomes large.
An important implication of our result is that Deutsch's model
turns quantum theory into a classical theory, in the sense
that each density operator becomes a distinct, distinguishable
point in a classical phase space.

\begin{figure}[ptb]
\begin{center}
\includegraphics[
width=3.0399in
]{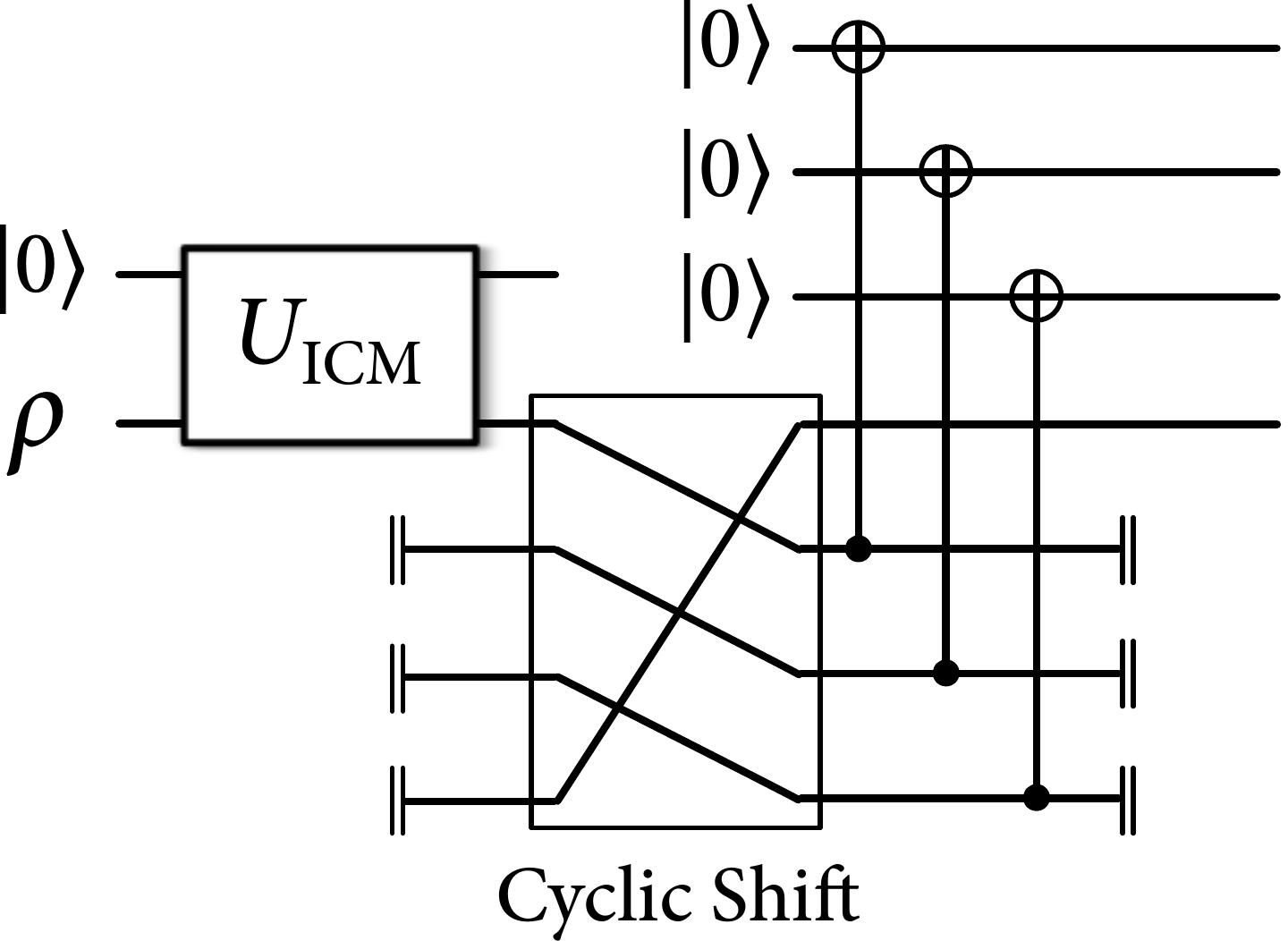}
\end{center}
\caption{Example circuit for cloning using $N=3$ CTC systems.
An unknown state $\rho$ is fed into a unitary $U_{\text{ICM}}$,
whose effect is to implement an informationally-complete measurement with
operators $\{M_x\}$, such that $M_x \geq 0$ and $\sum_x M_x = I$.
The resulting state $\omega = \sum_x \text{Tr} \{M_x \rho\} \vert x \rangle \langle x \vert$
is combined with $N$ CTC systems and cyclically permuted with them. (For each CTC system,
the past mouth of its wormhole on the left, indicated by vertical double lines,
is identified with its future mouth on the right.) Finally,
modular addition circuits
(depicted here as CNOT gates) ``read out'' $N$ copies of the state $\omega$, from which
we can estimate the original state $\rho$ to arbitrarily good accuracy as the number
$N$ of CTC systems becomes large (of course, one would require $N$ to be much
larger than three). The main text provides details of why this approach
works for D-CTC systems.} 
\label{fig:cloner}%
\end{figure}

One can quickly grasp the main idea behind our construction by taking a glance
at the circuit in Figure~\ref{fig:cloner}. The first step is to perform
an informationally-complete measurement on the incoming state $\rho_{S}$. Such
a measurement is well known in quantum information theory
\cite{P77,B91,RBSC04}---the probabilities of the outcomes are in one-to-one
correspondence with a classical density operator description of the quantum
state. (That is, if one knew these probabilities, or could estimate them from
performing this kind of measurement on many copies of the given state, then one
could construct a classical description of the state.) Let $\omega$ denote the
state resulting from the measurement:%
\begin{equation}
\rho \to \sum_{x=0}^{d-1} \text{Tr}\left\{  M_{x}\rho\right\}  \left\vert
x\right\rangle \left\langle x\right\vert \equiv \omega, \label{eq:measurement-map}
\end{equation}
where each $M_{x}$ is an element of the informationally-complete measurement
(so that $M_{x}\geq0$ for all $x$ and $\sum_{x}M_{x}=I$), $d$ is the number of
possible measurement outcomes, and $\left\{\left\vert x\right\rangle \right\}$
is the standard computational basis.

Next, we feed the state $\omega$\ into a circuit that cyclically permutes it
with $N$ CTC\ systems that each have the same dimension as $\omega$. Such an
operation on its own (after tracing over all systems except for the
$N$\ CTC\ systems) has as its unique fixed point the state $\omega^{\otimes N}$,
so that, in some sense, the cyclic shift produces $N$ ``temporary''\ clones.

Finally, we copy the value of $x$ from each of the $N$~CTC\ systems to one of
a set of ancillary systems in order to ``read out''\ $N$ copies of the state $\omega$.
In Figure~\ref{fig:cloner} we have depicted this operation as a sequence of controlled-not (CNOT)
gates, but in fact it will generally be a higher-dimensional analogue of a CNOT, like a modular
addition circuit:
\begin{equation}
\ket{x}\ket{y} \rightarrow U(\ket{x}\ket{y}) = \ket{x}\ket{(x+y)\bmod d} .
\label{eq:modular_addition}
\end{equation}
The fixed point of the overall circuit, after tracing over all systems
except for the $N$ CTC\ systems, is still $\omega^{\otimes N}$, because these
modular addition gates do not cause any disturbance to the CTC systems. As a
result, the reduced state on the $N$ ancillas is equal to $\omega^{\otimes N}$, and
we can then estimate the eigenvalues of $\omega$ simply by counting
frequencies---the estimates become better and better as $N$ becomes larger due to
the law of large numbers. Since these eigenvalues result from an
informationally-complete measurement, we can construct a classical description of
the state $\rho$ and produce as many approximate copies of it as we wish.

We now develop this argument in more detail. We first show how to produce
perfect clones of a quantum state that is diagonal in a known eigenbasis.
Suppose
that the initial state of the system and the CTC\ is as follows:%
\begin{equation}
\rho_{S}\otimes\sigma_{C},
\end{equation}
where $S$ is a $d$-dimensional system and $C$ consists of $N$ $d$-dimensional
systems. Furthermore, let $\rho_{S}$ have the following spectral
decomposition:%
\begin{equation}
\rho_{S}=\sum_{x}p_{X}\left(  x\right)  \left\vert x\right\rangle \left\langle
x\right\vert _{S},
\end{equation}
where $p_{X}\left(  x\right)  $ is a probability distribution and $\left\{
\left\vert x\right\rangle _{S}\right\}  $ is some orthonormal basis. The first
operation is to perform a cyclic shift by one to the right of all $N+1$
systems, i.e., the following unitary operation:%
\begin{multline}
\left\vert x_{1}\right\rangle _{S}\otimes\left\vert x_{2}\right\rangle
_{C_{1}}\otimes\left\vert x_{3}\right\rangle _{C_{2}}\otimes\cdots
\otimes\left\vert x_{N+1}\right\rangle _{C_{N}}\\
\rightarrow\left\vert x_{N+1}\right\rangle _{S}\otimes\left\vert
x_{1}\right\rangle _{C_{1}}\otimes\left\vert x_{2}\right\rangle _{C_{2}%
}\otimes\cdots\otimes\left\vert x_{N}\right\rangle _{C_{N}},
\end{multline}
where we have broken up the system $C$ into $N$ parts as $C_{1}\cdots C_{N}$.
One can then easily prove that, if this is the only interaction, the
self-consistent and unique solution is for the CTC\ systems to be in the state
$\rho^{\otimes N}$. Indeed, we can do so by demonstrating that $\rho^{\otimes
N}$ is the unique fixed point of the above map. For simplicity, let us
initialize the state of the CTC\ system so that it is maximally mixed, and so
that the overall state is%
\begin{equation}
\rho_{S}\otimes\pi_{C_{1}}\otimes\pi_{C_{2}}\otimes\cdots\otimes\pi_{C_{N}},
\end{equation}
where $\pi$ is the maximally mixed qudit state. After a cyclic shift, the
state becomes
\begin{equation}
\pi_{S}\otimes\rho_{C_{1}}\otimes\pi_{C_{2}}\otimes\cdots\otimes\pi_{C_{N}}.
\end{equation}
Tracing over the system $S$ gives
\begin{equation}
\rho_{C_{1}}\otimes\pi_{C_{2}}\otimes\cdots\otimes\pi_{C_{N}}.
\end{equation}
This becomes the initial state of the CTC\ for the next application of the
map, so that the overall state is now%
\begin{equation}
\rho_{S}\otimes\rho_{C_{1}}\otimes\pi_{C_{2}}\otimes\cdots\otimes\pi_{C_{N}}.
\end{equation}
Applying the cyclic shift again gives
$
\pi_{S}\otimes\rho_{C_{1}}\otimes\rho_{C_{2}}\otimes\cdots\otimes\pi_{C_{N}},
$
so that the reduced state is
$
\rho_{C_{1}}\otimes\rho_{C_{2}}\otimes\cdots\otimes\pi_{C_{N}}.
$
It is then clear that applying the above procedure $N$ times in total
gives the following state for the CTC%
\begin{equation}
\rho_{C_{1}}\otimes\rho_{C_{2}}\otimes\cdots\otimes\rho_{C_{N}}%
,\label{eq:cyclic-shift-fixed-point}%
\end{equation}
and further applications will not change anything, so that this is the fixed
point of the CTC. (In fact, by taking an arbitrary initial state for the CTC,
we can easily see by a similar procedure that the state in
(\ref{eq:cyclic-shift-fixed-point}) will be the unique fixed point after
applying the map $N$ times.)

Now, this procedure already produces $N$ temporary clones
of the initial state, and one might claim
that this circuit on its own is a cloner. However, the $N$ clones in the
CTC\ systems are not available after these systems enter the future mouth of
the wormhole, so that this cloner is not particularly useful. We would like
to have a circuit for which the clones are available after the CTC systems are no
longer in existence.

Since we have assumed for now that we know the eigenbasis of the incoming state, there
is a simple modification of the above circuit that will allow for cloning
it. Consider again performing the circuit given above. As we
showed, the fixed point solution for the CTC\ is $\rho^{\otimes N}$. What we
can do after the cyclic shift is to copy the value of $x$ from the
$N$ CTC\ systems to $N$ $d$-dimensional ancilla states initialized to the
state $\left\vert 0\right\rangle$, by using a modular addition circuit.
These circuits perform the unitary in (\ref{eq:modular_addition})
in the eigenbasis of the incoming system; they therefore
cause no disturbance to the CTC\ systems, and the fixed point
solution for the state of the CTC\ is still $\rho^{\otimes N}$. Furthermore,
the marginal state on the ancillas and the original system is $\rho^{\otimes
N+1}$, so that we have successfully produced $N$ clones of the state of the
incoming system, in the case where its eigenbasis is known. (If the eigenbasis is not known,
then one can easily check that our circuit will decohere the incoming state $\rho$ in the basis
in which the modular addition circuits are specified and produce $N$ perfect copies of
the decohered state.)

The above circuit allows for {\it perfect} cloning of quantum states in a known
eigenbasis.  A particular preprocessing of an arbitrary incoming state will
allow us to produce {\it approximate} clones whose fidelity with the incoming
state becomes arbitrarily high in the limit where the number $N$ of
CTC\ systems becomes large.
Let $\rho$ denote the density operator of the input state.
We can perform a measurement map of the form in (\ref{eq:measurement-map}) on the incoming state.
Such a map is a
CPTP\ map, so that we can perform it by first appending an ancilla of
sufficient size, acting with a unitary on the joint system, and tracing out
the ancilla. We should be sure to choose the measurement map to be informationally
complete, such that the outcome probabilities are in one-to-one correspondence with
the parameters of the density operator.

The procedure for approximate cloning is as follows:

\begin{enumerate}
\item On the incoming state, perform the measurement map specified by
(\ref{eq:measurement-map}).

\item Append the $N$\ CTC\ systems to this state and send the $N+1$ systems
through the cyclic shift circuit, followed by $N$ CNOT gates from
the CTC\ systems to $N$ ancilla systems.

\item The resulting state after the CTC\ expires is $\omega^{\otimes N+1}$
(with $\omega$ defined in (\ref{eq:measurement-map})).

\item Perform measurements in the basis $\left\{  \left\vert x\right\rangle
\right\}  $ to estimate the distribution Tr$\left\{  M_{x}\rho\right\}  $\ to
arbitrarily good accuracy (with $N$ large).

\item Based on the estimate, produce as many approximate clones of $\rho$ as desired.
\end{enumerate}

In Step~4, we can argue that the estimate becomes arbitrarily good as $N$
becomes large, due to the law of large numbers. In particular, Hoeffding's
bound states that the probability for the empirical frequencies to deviate
from their true values by more than any constant $\delta>0$ is bounded from
above as $2\exp\left\{  -2N\delta^{2}\right\}  $ \cite{H63}, so that this probability
rapidly converges to zero as the number $N$ of CTC\ systems increases.
The 
number of CTC systems scales well
 with the desired accuracy for cloning (and the number of gates is linear
 in the number of CTC systems)---to have an estimation error
no larger than some constant $\varepsilon > 0$ requires a number of gates
no larger than $O( \log(1/\varepsilon))$.

A slight modification of the above protocol would be to
avoid tracing over the ancilla after 
performing the unitary corresponding
to the measurement map. The state resulting from the unitary is
$
\sum_{x,y} (M_x \rho M_y^\dag) _E \otimes \vert x\rangle \langle y\vert_B
$
if
the input state is $\rho$,
where we have labeled the environment as $E$ and the output as $B$.
We could then append $N$ CTC systems, labeled as $E_1 B_1 \cdots E_N B_N$
that are each the same dimension
as the composite system $EB$.
After that, we would perform a cyclic shift of the $E_i B_i$ systems, followed by
a CNOT gate from each $B_i$ system to an external ancilla. It is straightforward
to show that the fixed-point solution of the CTC systems $E_1 B_1 \cdots E_N B_N$ is then
$
\left( \sum_x M_x \rho M_x^\dag
\otimes \vert x\rangle \langle x \vert \right)^{\otimes N} .
$
That is, the effect of the
CNOT gates is to decohere the $B_i$ systems. The CNOT
gates will then read out many copies of the state
$\sum_x \text{Tr} \{M_x \rho \}
\vert x \rangle \langle x \vert$ to the external ancillas, from which
we can estimate the input state $\rho$ as before.
An advantage of this approach is that this modified circuit avoids
potential interpretational issues with the initial measurement map. That is,
one might claim that the measurement map in (\ref{eq:measurement-map}) actually
``collapses'' the state $\rho$ to one of the states $\vert x \rangle$
with probability $\text{Tr} \{M_x \rho \}$
and the resulting circuit merely copies the given state $\vert x \rangle$
many times, providing no advantage for cloning over an ordinary quantum circuit.
However, by having all evolutions be unitary, it is clear that the modified circuit 
 avoids this interpretational problem.

By a well-known argument \cite{D82}, the ability to clone implies
the ability to signal superluminally, so that this is the case for our cloner here
(assuming the usual description of quantum measurements).
Our results imply more generally
that every continuous, but otherwise arbitrarily non-linear map $f$
from states to states can be implemented to arbitrary accuracy with Deutschian CTCs.
This follows because we can estimate the incoming state $\rho$ to arbitrary accuracy
and then prepare $f(\rho)$ at will.

{\it Discussion}---An ``open timelike curve'' is one in which a quantum system enters the future
mouth of a wormhole and emerges from the past mouth of the wormhole without ever interacting with itself along the way \cite{PhysRevLett.110.060501}.  Our circuit in Figure~\ref{fig:cloner} indicates that we are very close to implementing quantum state cloning using only an open timelike curve.  If the modular addition circuits were not present, then this approach would indeed be just an open timelike curve. We say that we are ``very close'' because in our setup, the modular addition circuits do not disturb the state of the CTC systems, so one might be tempted to expand the definition of an open timelike curve to allow for such non-disturbing interactions.

One might question the method above by taking an adversarial approach to
quantum state cloning as was done with quantum state discrimination
in \cite{PhysRevLett.103.170502}. In such an
adversarial model as described in \cite{PhysRevLett.103.170502},
an adversary would prepare a labeled mixture of states of the form
$\sum_x p(x) \vert x \rangle \langle x \vert \otimes \rho_x$, feed in the second system
to the cloner,
and demand that the output state of the composite system be
$\sum_x p(x) \vert x \rangle \langle x \vert \otimes \tilde{\rho}_x \otimes \tilde{\rho}_x,$
where $\tilde{\rho}_x$ is a good approximation to $\rho$. Our approach
will not satisfy this demand but instead outputs an approximate copy of the
average state $\sum_x p(x)  \rho_x$ because Deutsch's criterion in (\ref{eq:fixed-point})
stipulates that the fixed point is computed with respect to the reduced state of the system 
entering the CTC device.
However, such behavior is to be expected,
since quantum mechanics in Deutsch's model is no longer linear, so that the action of
a map on a mixture of states is not equal to the mixture of states resulting from the map
acting on each state. Some authors have argued that it
is not sensible to represent ensembles as labeled mixtures when we are dealing
with a non-linear theory \cite{CM10}.  Labeled mixtures are in one-to-one correspondence with
ensembles in standard quantum mechanics, but this correspondence
breaks down in a non-linear theory. One might also argue that all of this points
to the Deutsch model itself being incomplete \cite{CMP12}.
Regardless, what we have shown in this 
paper is that if a quantum state $\rho$ is presented to a device that behaves according to the
prescription of Deutsch's model, then it is possible to produce an arbitrary number
of very good approximate clones of $\rho$.

Our results imply that, in a particular sense,
Deutsch's model is actually a classical model
for CTCs rather than a quantum model. That is,
quantum theory supplemented with Deutsch's prescription
for CTCs seems to require an interpretation as a
classical probability theory on the space of
density operators. This in turn leaves open the
question of whether other descriptions of CTCs
might retain more of the distinctive features of quantum theory.
This feature of Deutsch's model originates from the way that it combines quantum
features (density operators and unitary evolutions) with non-quantum
ones (non-linear evolution) in an {\it ad~hoc} way.
In the framework of generalized probabilistic theories \cite{H01,B07,CMP10}, there is a
basic result stating that, if states and evolutions are defined
operationally, then the evolution must be linear in the
state \footnote{``Linear'' is meant
here in the operational sense: a state is a linear combination of a
set of states if the outcome probabilities for that state are linear
combinations of the outcome probabilities of those states, for every
possible measurement; relative to this operational notion of
linearity, a physical transformation must send a linear combination of
input states to a linear combination of output states.}.
The fact that Deutsch's model allows for evolutions that are
non-linear in the density operator implies that the set of all density
operators corresponds to linearly independent states at the operational
level. Our result strengthens this observation for the case of Deutsch's model,
showing
that the states corresponding to density operators are not only
linearly independent, but even perfectly distinguishable.

{\it Open Questions}---It might be considered somewhat unsatisfactory
that we obtain only an arbitrarily good approximation of cloning. The limit
achieving perfect cloning requires
taking the size of both the CTC system and the ancillary system to infinity.
This seems to be necessary if we would like to read out the parameters of
the density operator as we have done here, but this is less clear if all we
desire is to have two clones of the incoming state.
So an open question to consider going forward from here is if there
exists an exact $1 \to 2$ CTC-assisted cloner
that requires only a finite external system but with a potentially infinite
internal CTC system.

We are grateful to Patrick Hayden and John Preskill for insightful discussions
and especially to the anonymous referees for insightful comments and additions.
MMW is grateful for the hospitality of
the Ming Hsieh Department of Electrical Engineering at the University of Southern California and
F\'{\i}sica Te\`{o}rica: Informaci\'{o} i Fenomens Qu\`{a}ntics at the
Universitat Aut\`{o}noma de Barcelona and for support for research visits during April
and May 2013.
AW acknowledges financial
support from the Spanish MINECO (project FIS2008-01236), the European
Commission (STREP ``QCS''), the ERC (Advanced Grant ``IRQUAT'') and a
Philip Leverhulme Prize.

\bibliographystyle{unsrt}
\bibliography{Ref}

\end{document}